\documentclass[a4paper]{article}  
\usepackage{indentfirst}  
\usepackage{bm}    
\usepackage{graphicx}  
\usepackage{url}
\usepackage{epstopdf}
\usepackage{fancyhdr}

\usepackage{geometry}
\usepackage{color}
\newcommand{\kinectTM}{Kinect\texttrademark}
\newcommand{\kinectTMS}{\kinectTM~}

\newcommand{\sectTech}{2}
\newcommand{\sectWdyn}{3}
\newcommand{\sectAnalysis}{3.1}
\newcommand{\sectConcl}{4}

\newcommand{\figSketchStat}{1}
\newcommand{\figCompositionImg}{2}
\newcommand{\figSampleTrajs}{3}
\newcommand{\figDensityHistory}{4}
\newcommand{\figBasicFdiagr}{5}
\newcommand{\figVelRatios}{6}
\newcommand{\figFluxFdiagr}{7}

\begin{document}    
\begin{center}
\LARGE\bf Continuous measurements of real-life bidirectional pedestrian flows on a wide walkway
\end{center}

\begin{center}
\rm Alessandro Corbetta$^1$, \ \ Jasper Meeusen$^2$, \ \ Chung-min Lee$^3$ \ and \ Federico Toschi$^{4}$
\end{center}

\begin{center}
\begin{small} \sl
${}^{\rm 1,2,4}$ Eindhoven University of Technology, 5600 MB, Eindhoven, The Netherlands \\   
a.corbetta@tue.nl; j.a.meeusen@student.tue.nl; f.toschi@tue.nl \\
${}^{\rm 3}$ California State University Long Beach, Long Beach, CA 90840, USA \\   
chung-min.lee@csulb.edu \\
${}^{\rm 4}$ CNR-IAC, I-00185,  Rome, Italy \\   
\end{small}
\end{center}
\vspace*{2mm}

\begin{center}
\begin{minipage}{15.5cm}
\parindent 20pt\small
\noindent\textbf{Abstract - } 
Employing partially overlapping overhead \kinectTMS sensors and automatic pedestrian tracking algorithms we recorded  the crowd traffic in a rectilinear section of the main walkway of Eindhoven train station on a 24/7 basis. Beside giving access to the train platforms (it passes underneath the railways), the walkway plays an important connection role in the city. Several crowding scenarios occur during the day, including high- and low-density dynamics in uni- and bi-directional regimes. In this paper we discuss our recording technique and we illustrate preliminary data analyses. Via fundamental diagrams-like representations we report  pedestrian velocities and fluxes vs. pedestrian density. Considering the density range $0$ - $1.1\,$ped/m$^2$, we find that at densities lower than $0.8\,$ped/m$^2$ pedestrians in unidirectional flows walk faster than in bidirectional regimes. 
On the opposite, velocities  and fluxes for even bidirectional flows are higher above $0.8\,$ped/m$^2$.
\end{minipage}
\end{center}

\begin{center}
\begin{minipage}{15.5cm}
\begin{minipage}[t]{2.3cm}{\bf Keywords:}\end{minipage}
\begin{minipage}[t]{13.1cm}
  Pedestrian Dynamics; High Statistics Measurements; Continuous Pedestrian Tracking; Microsoft \kinectTM; Bidirectional Pedestrian Flows; Fundamental Diagrams
\end{minipage}\par\vglue8pt
\end{minipage}
\end{center}

\section{Introduction}  
The importance of a good understanding and an accurate modeling  of the behavior of pedestrian crowds, e.g. to  support design or operation decisions in civil infrastructure or to explain the complex features in pedestrian motion, demands for  accurate and extensive experimental data. 

Improvements in the measurement techniques over time enabled   data collection at the single   (i.e. ``microscopic'') pedestrian level,  by means of automatic head tracking~\cite{DBLP:journals/ijon/BoltesS13,seer2014kinects}. Measurements in real-life are nowadays also a possibility: overhead 3D depth maps of pedestrian crowds (as those provided by Microsoft \kinectTMS sensors~\cite{Kinect} employed here, cf. Section~\sectTech \ for technical details) allow a reliable and privacy-respectful  pedestrian tracking~\cite{seer2014kinects}. This enables potentially unlimited  data collection out-of-laboratories~\cite{Brscic201477,corbetta2014TRP,corbetta2016eulerian}, involving pedestrians behaving naturally and unaware  of taking part to recordings for a  scientific experiment. This can lead to arbitrary accuracy and statistical resolution including tails and rare events~\cite{corbetta2016multiscale}. 

The long-time debated influence of bidirectionality on pedestrian flows, namely the modification of ``macroscopic'' fluxes and transport properties from  unidirectional flows to bidirectional counterflows, received increasing insights from technology development. However, in more than $40$ years of studies, no general consensus has yet been reached. In comparison to unidirectional streams, counterflows have been reported to yield fluxes that are negligibly different~\cite{weid,fruin1987BOOK}, lower~\cite{pushkarev1975capacity,lam2002study,zhang2012ordering} or, because of the self-organization of the stream, higher~\cite{kretz,helbing2005self,corbettaTGF15}.
In all these cases, fundamental diagrams~\cite{weid},  macroscopic crowd density--velocity or density--flux relations, are typically employed in the comparison.

To inquire the properties of uni- and bi-directional pedestrian flows in real-life and with high statistic resolution we established a 24/7 pedestrian tracking site at Eindhoven train station (Eindhoven, The Netherlands). Extending our previous \kinectTMS infrastructure~\cite{corbetta2014TRP,corbetta2016eulerian,corbettaTGF15,corbetta2015MBE} for continuous coverage of large spaces we analyzed a full section of the main walkway, which is relatively highly crowded over the day. In this paper we discuss pedestrian fundamental diagrams, also in dependence on the ``extent of bidirectionality'' of the flow, i.e. we compare the unidirectional regime with even and uneven  bidirectional traffic. Thanks to the large volume of data collected (tens of millions of frames in total), we are able to report fundamental diagrams in terms of resolved probability distribution functions (pdf), possibly in dependence of the direction ratio (formally defined hereafter).

The paper is structured as follows: in Section \sectTech \  we provide an overview of our measurement location and of our acquisition approach via multiple overhead Microsoft \kinectTMS sensors. In Section \sectWdyn \ we exemplify the daily pedestrian dynamics in the station walkway and define the physical observables that we analyze in Section \sectAnalysis. The discussion in Section \sectConcl \ closes the paper.

\vspace*{4mm}
\centerline{\includegraphics[width=.4\textwidth]{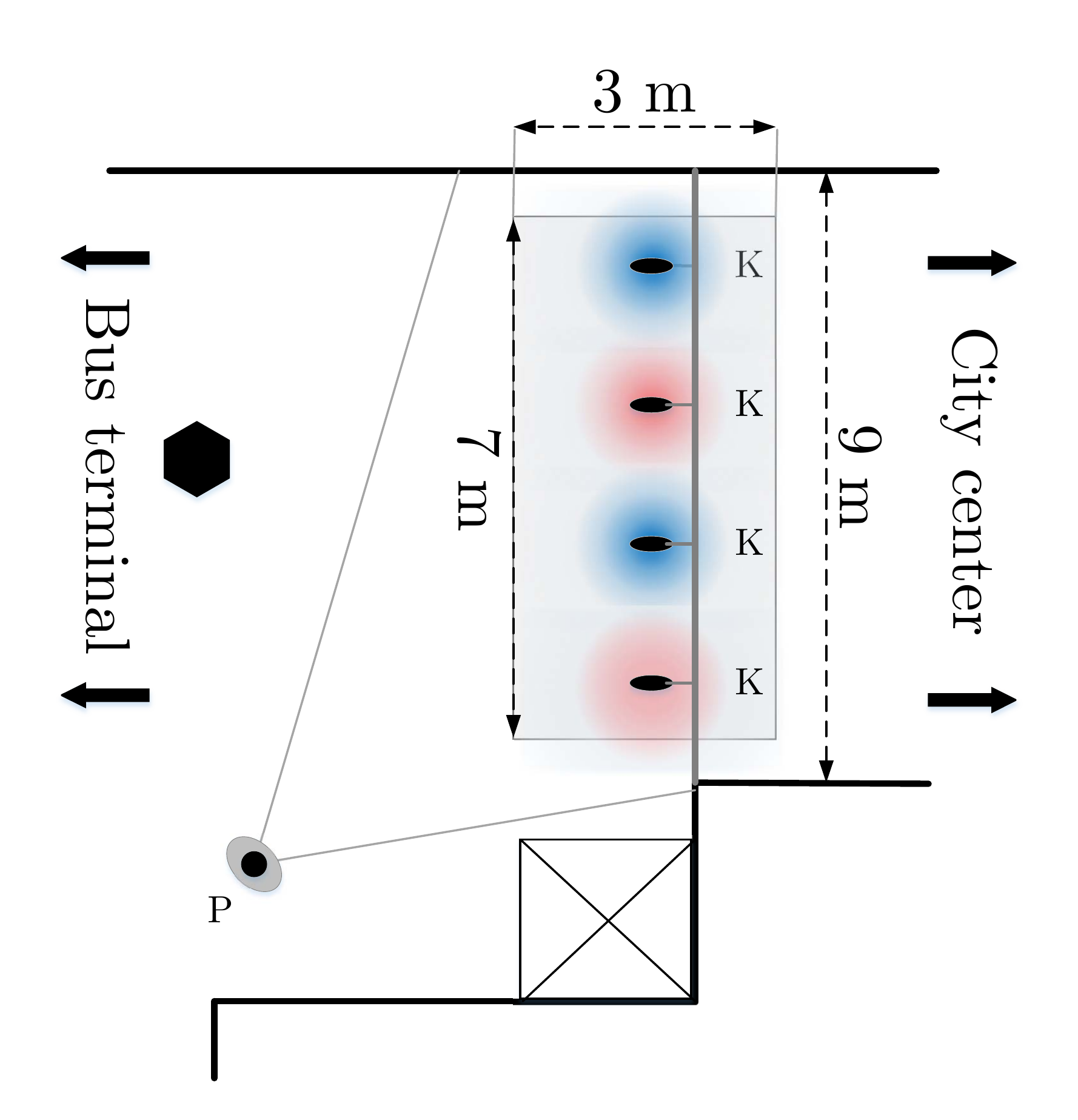}
\includegraphics[width=.5\textwidth]{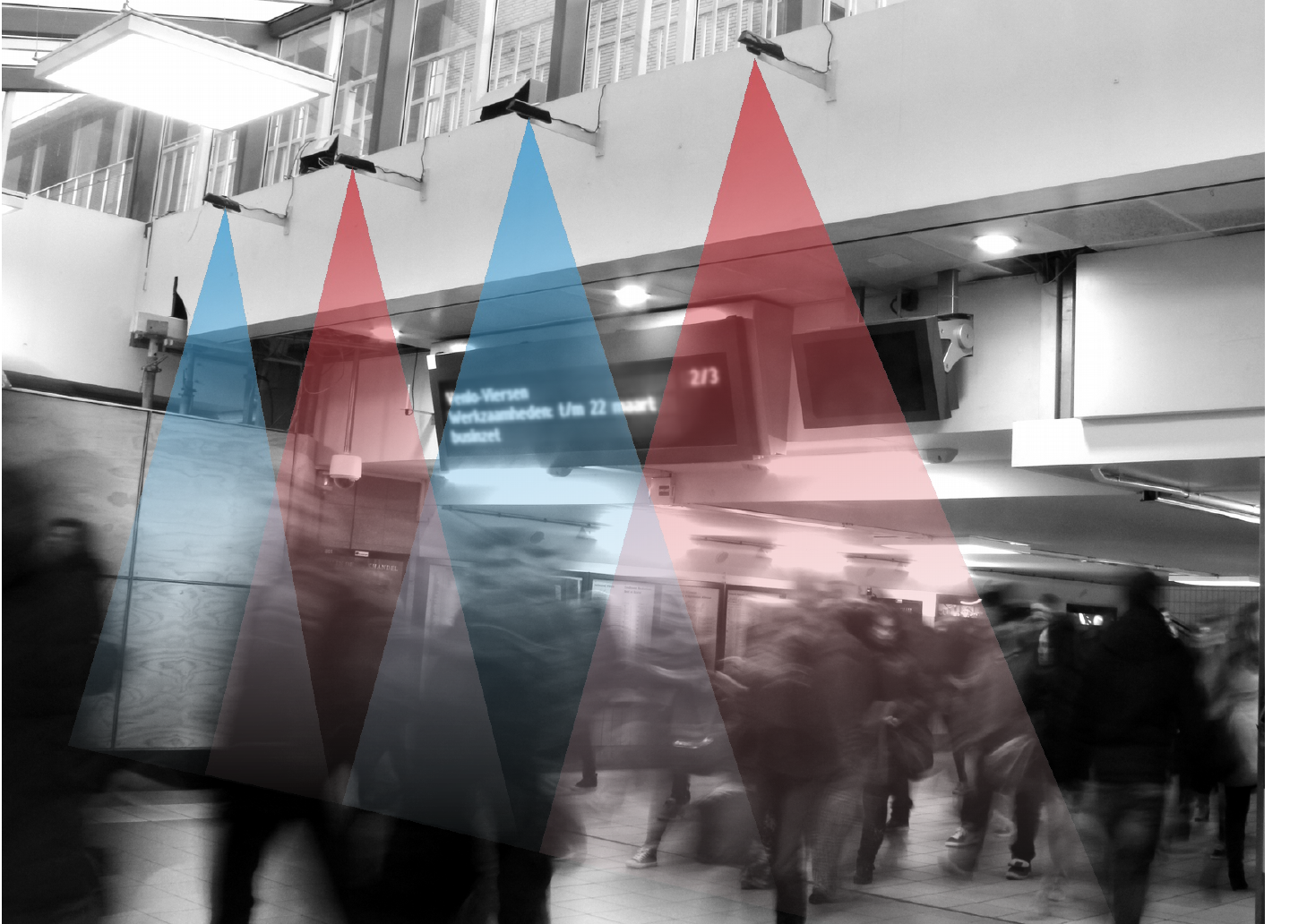}
}
\begin{center}
  \parbox{15.5cm}{\small\center{\bf Fig.\figSketchStat.} (Left) Planar view of the north entrance of Eindhoven Train Station (in the configuration before the mid-2015 renovation). Pedestrians arriving from the main bus terminal enter the building. A tunnel $70\,$m long departs from the entrance hall and connects to the city center side of the station.  Train platforms (above) can be reached via staircases along the tunnel. A large pillar splits the pedestrian flow at  the entrance.  We recorded the initial part of the tunnel (grey area) via four overhead \kinectTMS sensors (\textit{K}) in partial overlap. (Right) Live view of the typical crowd flow recorded   as seen by observer \textit{P} (approximately 8:30 AM, March 2015). The height ($4\,$m) and mutual distance of the sensors ensures continuous spatial coverage of the area from the ground to about $2\,$m up from it.}
\end{center}
\vspace*{4mm}

\section{Real-life continuous measurements}\label{sect:tech}
Our pedestrian tracking campaign took place in the period September 2014 -- April 2015 and considered the entrance area of the main walkway of the train station. The walkway, a large tunnel, beside providing the access to the track platforms via flights of stairs (outside our tracking area) serves also as a major pedestrian connection from the north to the south side of the city (cf. Fig.~\figSketchStat). Our raw data recordings employed an array of four overhead and aligned Microsoft \kinectTMS depth sensors (placed as in Fig.~\figSketchStat) allowing exhaustive spatial coverage of the tunnel entrance. Overhead \kinectTM-based data proved to allow reliable automatic pedestrian positioning for tracking~\cite{seer2014kinects,corbetta2014TRP,brscic2013person}. In fact, overhead vision avoids the challenge of resolving mutual occlusions,  and depth fields  (the distance field between points and the sensor plane; see Fig.~\figCompositionImg\  and~\figSampleTrajs, where the depth field is reported, respectively, in colormap and in grayscale) ease the localization of individual heads and bodies (we refer to our previous work~\cite{corbetta2015MBE} and to~\cite{seer2014kinects} for the general segmentation approach).

\paragraph{Blending overhead depth maps} We aggregate the sensor streams at the depth image level, blending simultaneous depth maps into one large ``fusion'' map (see Fig.~\figCompositionImg). We then extract pedestrian positions from such maps to reconstruct trajectories via Particle Tracking Velocimetry algorithms~\cite{OpenPTV,willneff2003spatio}. This aspect marks a difference from other \kinectTM-based measurement campaigns~\cite{seer2014kinects,brscic2013person}, where pedestrian trajectories are matched and merged after processing the sensor streams individually. At the cost of performing  pedestrian detection algorithms on large depth maps (computationally expensive as algorithms~\cite{corbetta2015MBE} typically scale  more-than-quadratically in the number of pixels to process), we benefit from translational invariance over the whole maps and from resolved depth reconstruction at the  overlapping edge regions. The partial and skewed sensor view at the edges might in fact increase reconstruction and detection errors.

\vspace*{4mm}
\centerline{\includegraphics[width=.5\textwidth]{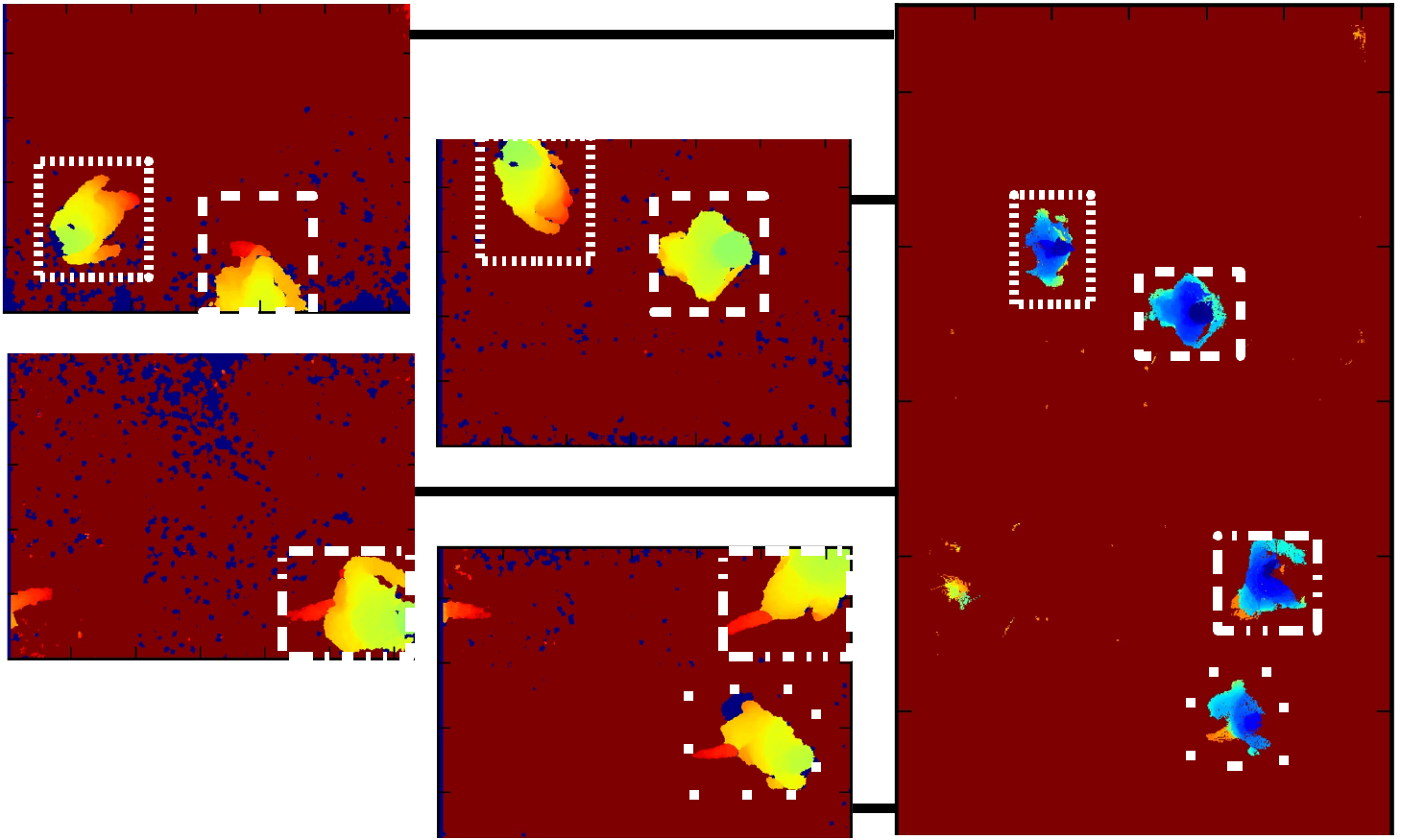}
\includegraphics[width=.4\textwidth]{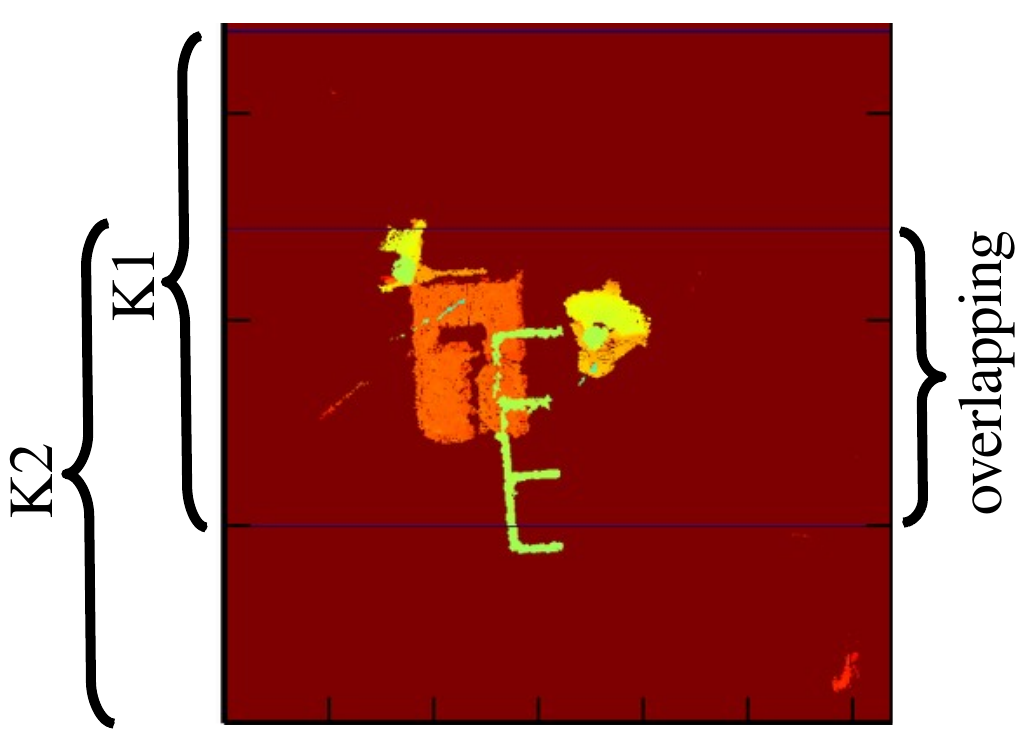}
}
\begin{center}
\parbox{15.5cm}{\small\center{\bf Fig.\figCompositionImg.} (Left) Raw depth signals from the four \kinectTMS sensors are superposed after an axonometric transformation of the depth map (cf. Eq.~(\ref{eq:coordinate-transf})). In the overlapping regions locally lower depth values (associated to points closer to the sensors) are retained. Bounding boxes in different line styles identify the same pedestrians in the frames. (Right) Calibration of sensors coordinates to resolve the overlapping region (case of two sensors: K1 and K2). We found the relative coordinates of each depth map (after axonometric transformation) by ensuring a  correct reconstruction of the ``calibration cart'' displayed (the cart is $1\,$m long and supports a beam  $1.5\,$m long  attached to transversal elements of known size). 
}
\end{center}
\vspace*{4mm}

We obtain  ``fusion'' depth maps from individual sensor data in a two steps process (we refer to the appendix in~\cite{corbetta2016multiscale} for further technical discussions):
\begin{enumerate}
\item \textit{Generation of aerial axonometric depth maps.} Depth maps come with the perspective view of the sensor. In other words, the sensor outputs a ``vertical'' aerial view just for pedestrians close to the sensor axis (i.e., right below the sensor location). Instead, pedestrians moving toward the edges of the view cone get a skewed image. Depth maps can be reprocessed to align points lying on the same vertical ray obtaining an aerial view. In this way depth maps of shapes moving on the horizontal plane are translation invariant. Let $(\theta,\rho,h)$ be the cylindrical coordinates (in the angle $\theta$) aligned with the camera axis ($\rho = 0$), normalized so that the view cone and the ground intersect at $\rho=1$ and $0 \leq h \leq 1$ spans (vertically) the depth (distance) from the sensor ($h=0$) to the ground ($h=1$). The transformation
\begin{equation}\label{eq:coordinate-transf}
(\theta,\rho,h) \mapsto (\theta,\rho h,h) 
\end{equation}
displaces each point to its vertical line (note that $\rho h \leq \rho$ and $\rho h = \rho$ only at the ground level, i.e. the ground level is invariant under this transformation). Extracting the lowest depth value (top-most) for each vertical line yields the final  aerial view.  
\item \textit{Juxtaposition of aerial depth maps.} We merge simultaneous aerial depth maps by juxtaposition/superposition according to the sensors overlap. We find superposition parameters via a calibration involving sliding beams of known size under the cameras and then reconstructing by a combination of the aerial views. Once more, the superposition procedure selects the lowest  depth value for each vertical line.
\end{enumerate}
In Fig.~\figCompositionImg\ we report an example of a ``fusion'' frame, along with a calibration frame for the overlapping of two \kinectTMS sensors next to each other in the array.  In Fig.~\figSampleTrajs\  we report  reconstructed pedestrian trajectories acquired at different density levels.

\vspace*{4mm}
 \centerline{\includegraphics[width=.275\textwidth]{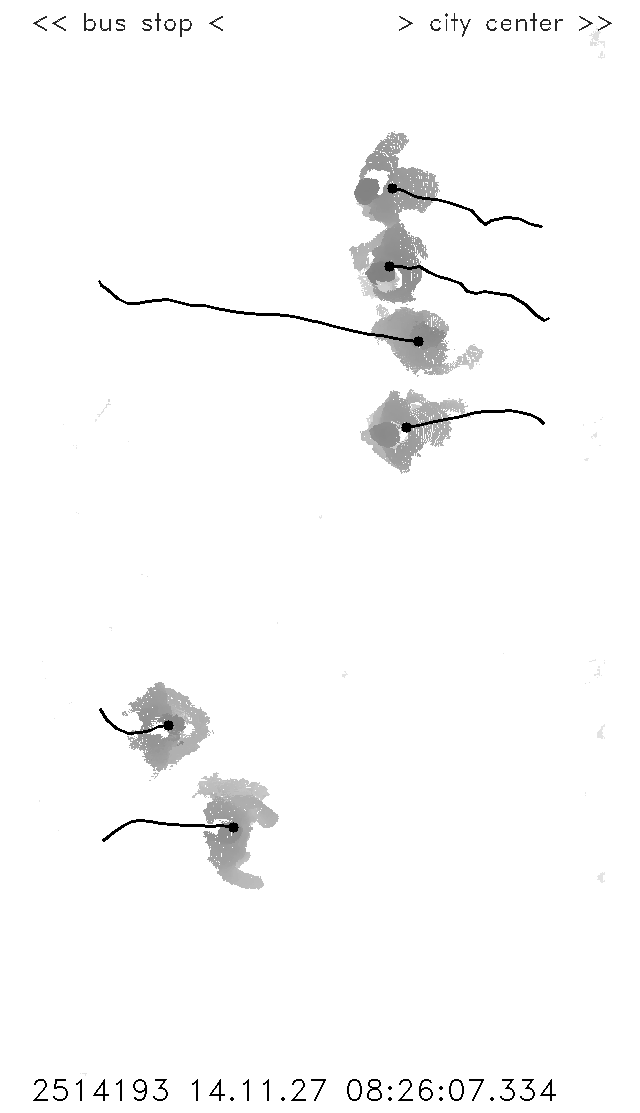}
 \includegraphics[width=.275\textwidth]{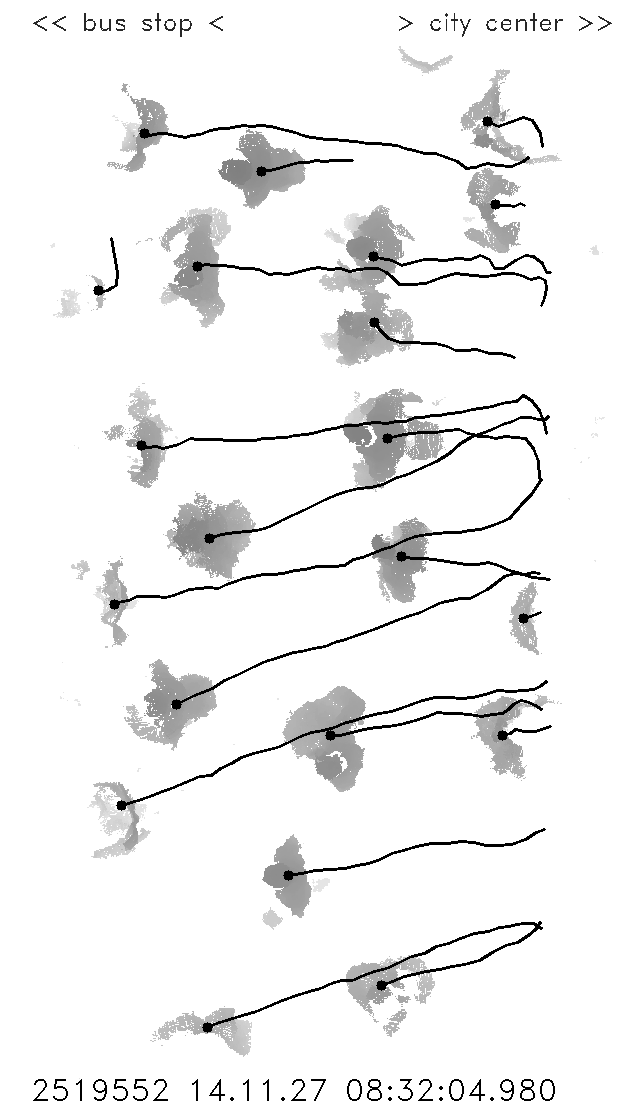}
\includegraphics[width=.275\textwidth]{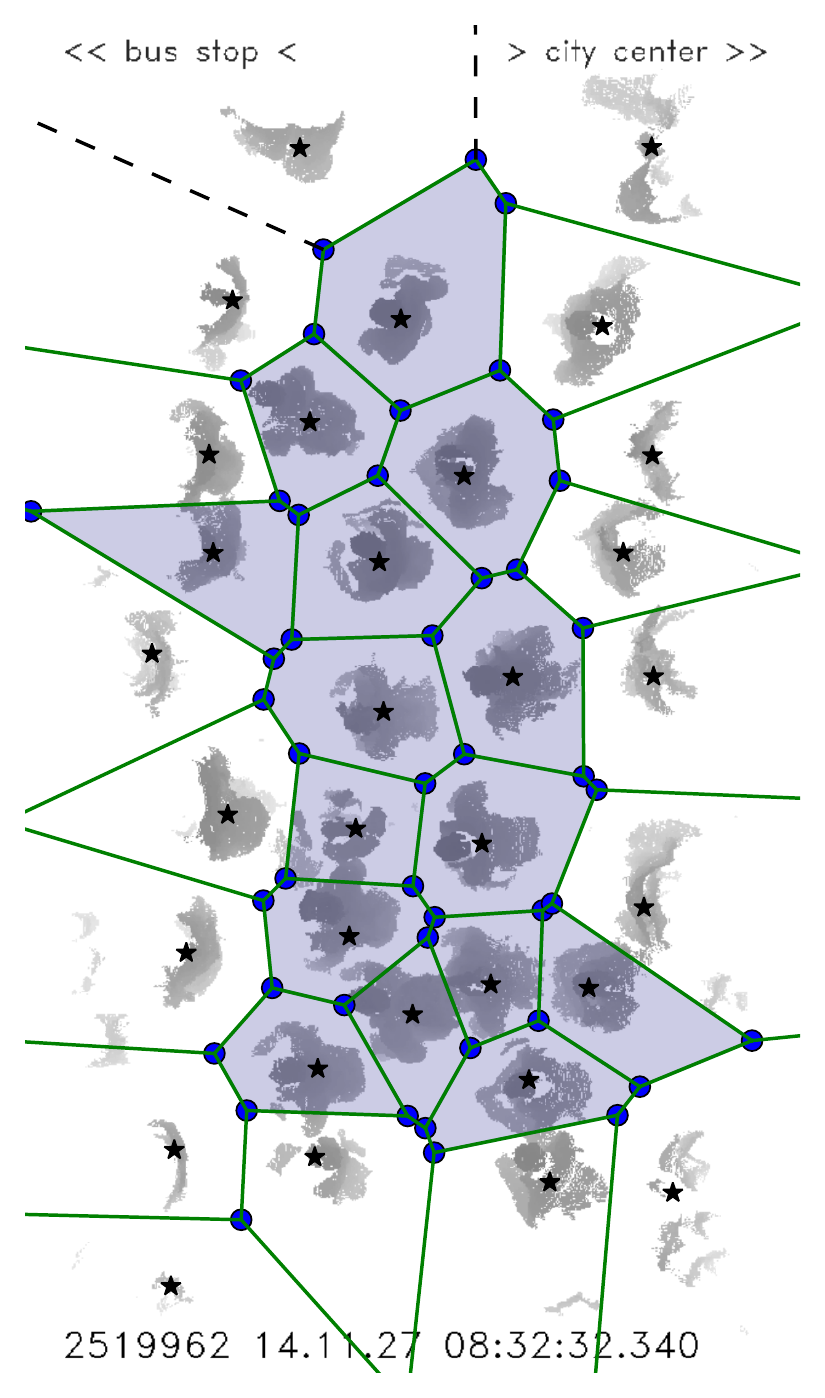}
}
\begin{center}
  \parbox{15.5cm}{\small\center{\bf Fig.\figSampleTrajs.} Pedestrian trajectories after automatic detection and tracking algorithms. At peak hour (around 8:30 AM, when commuters arrive) the  number of pedestrians in the observation region increases in few seconds from few up to more than forty pedestrians in a single frame. The figure shows three different traffic conditions: (left) low density, $6$ persons in even bidirectional flow ($r = 0.5$, cf. Eq.~(\ref{eq:direction})); (center and right) unidirectional flows at higher congestion with, respectively, $21$ and $32$ pedestrians. In the right figure we include a Vonoroi tessellation based on pedestrian positions to estimate the personal space/local density (cf. Section \sectAnalysis). Solid colored cells close within our observation region.}
\end{center}
\vspace*{4mm}

\vspace*{4mm}
\centerline{\includegraphics[width=1.\textwidth,trim=0cm -.5cm 0cm 0.cm]{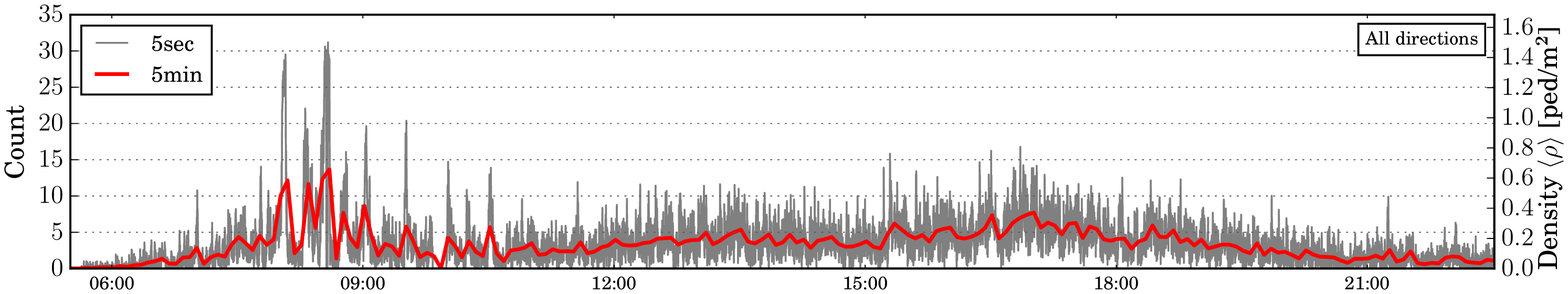}}
\centerline{\includegraphics[width=1.\textwidth]{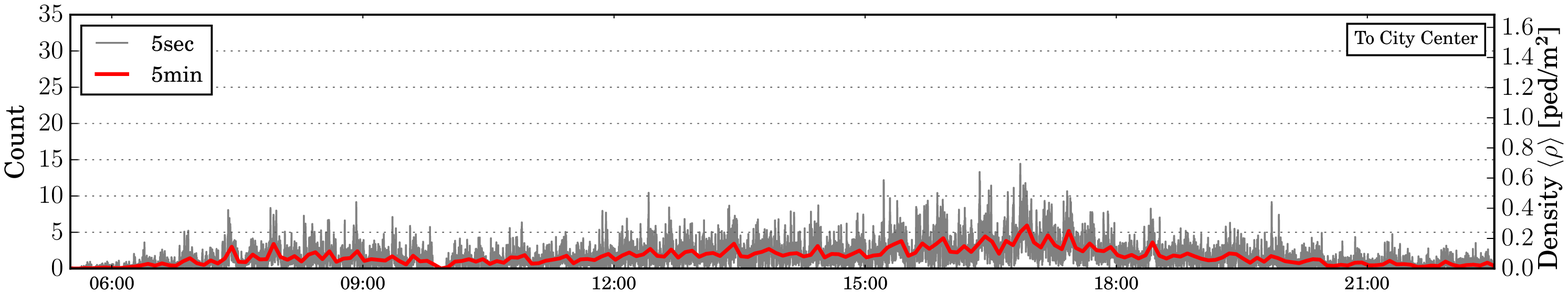}}
\centerline{\includegraphics[width=1.\textwidth]{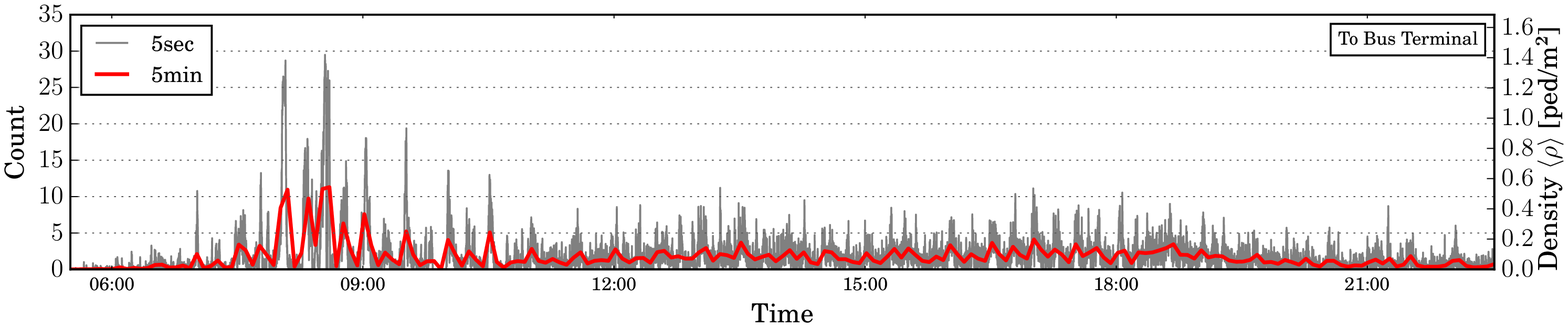}}
\begin{center} 
  \parbox{15.5cm}{\small\center{\bf Fig.\figDensityHistory.}
    Time history of pedestrian counts and density on the 27$^{th}$ of November 2014.  We report the overall count in the top plot; counts in dependence on the walking direction are reported in the middle and in the bottom plot (respectively, to the city center and to the bus terminal, cf. Fig.~\figSketchStat). The time line spans between 5:33 AM (arrival of the first train) till 11:30 PM. At 8:32 AM we measure the highest density as commuters' trains arrive shortly before from several cities at once. In the center plot (traffic from the bus terminal to the city center) we notice a load that is more uniform in time, with a small increase in density in the afternoon, when people leave the city by train. The pedestrian density (ratio between count and measurement surface area) is reported on the right hand side on the vertical axis.}
\end{center}
\vspace*{4mm}

\section{Pedestrian dynamics at Eindhoven train station}\label{sect:wdyn}
The daily pedestrian flow along the walkway, typically bidirectional, alternates between several quite different regimes. We recorded approximately $80,000$ trajectories during typical normal working days  (characteristic crossing time of the recording window is $3\,$s), with instantaneous counts and densities as in Fig.~\figDensityHistory.  The peak densities are measured at mornings around 8:30 following the arrival of commuters' trains. Most of pedestrians in these conditions walk towards the bus terminal. On the opposite, during afternoons, traffic is  more homogenous in time  and it is mostly due to pedestrians walking towards the city center/train platforms. This includes commuters on their way back home. During weekends we observe typically a $35\%$ reduction in traffic.

In the following we analyze  measurements from $55$ days (h$24$, acquisition rate was $15\,$fps) providing a total of $38\,$million frames showing  at least one pedestrian. On a frame-by-frame basis we compute the following quantities:
\begin{itemize}
\item \textit{Pedestrian average density/personal space.} Following~\cite{steffen2010methods} we evaluate the personal space around each pedestrian by means of a Voronoi tessellation based on the bodies centroids (cf. Fig.~\figSampleTrajs). The  area $A_p$ of the Voronoi cell associated to each pedestrian gives a measure of the personal space available. Its reciprocal $\rho_p = A_p^{-1}$ provides a microscopic notion for the local density.  Voronoi cells in the neighborhood of the  boundaries or at low densities remain open. In these cases we approximate the personal space as $A_{p,open} = (A_t - \sum_p A_p)/N_{open}$, where $A_t$ is the total recording area, $\sum_p A_p$ is the total area of the closed Voronoi cells and $N_{open}$ is the amount of pedestrians with open cell.  We denote the frame-wise average of such local densities as  $\langle\rho\rangle$.
\item \textit{Direction ratio.} To quantify the bidirectionality of the traffic flow, we compute the ratio between the total personal area for pedestrians going to the city center ($A_{city}$) and to the bus terminal ($A_{bus}$). We define
\begin{equation}\label{eq:direction}
r= 1 - \frac{\max(A_{city},A_{bus})}{A_T},
\end{equation}
which for $r=0$ (minimum) indicates a unidirectional flow, while for $r=1/2$ (maximum) indicates an evenly distributed bidirectional flow.
\item \textit{(Average) walking speed/Average flux.} In association with the average density $\langle\rho\rangle$, we describe the  frame-wise dynamics via the signed walking speed $v$ (one contribution per pedestrian per frame). We further consider frame-wise averages of the absolute walking speed, indicated by  $\langle |v| \rangle$, and of the absolute flux $\langle\rho\rangle\langle |v| \rangle$.
\end{itemize}

\subsection{Fundamental diagrams}
We characterize the pedestrian flows via functions of the average density $\langle\rho\rangle$, such as fundamental diagrams. Thanks to our high data volume (cf. Fig.~\figBasicFdiagr\,(left)), we possibly report full density-dependent pdfs of the measurements (after  binning the $\langle\rho\rangle$ range $[0,1.6]\,$ped/m$^2$ into $32$ equal intervals). For ease of visualization, we plot the contour levels of these pdfs across the density bins.  
We further include average and standard deviation plots for readability.   In Fig.~\figBasicFdiagr\,(right) we plot the walking speed pdf vs. the density. Beside an expected decreasing trend, as the density increases, we  notice a reduction in the speed variability. At low densities more dynamics are possible and it is easier for pedestrians to even stand still or to run. As the density increases, this freedom reduces since the motion becomes more constrained, thus it is natural to expect that the speed variance should decrease.

\vspace*{4mm}
\centerline{\includegraphics[width=.85\textwidth,trim=0cm 0cm 0cm 0cm]{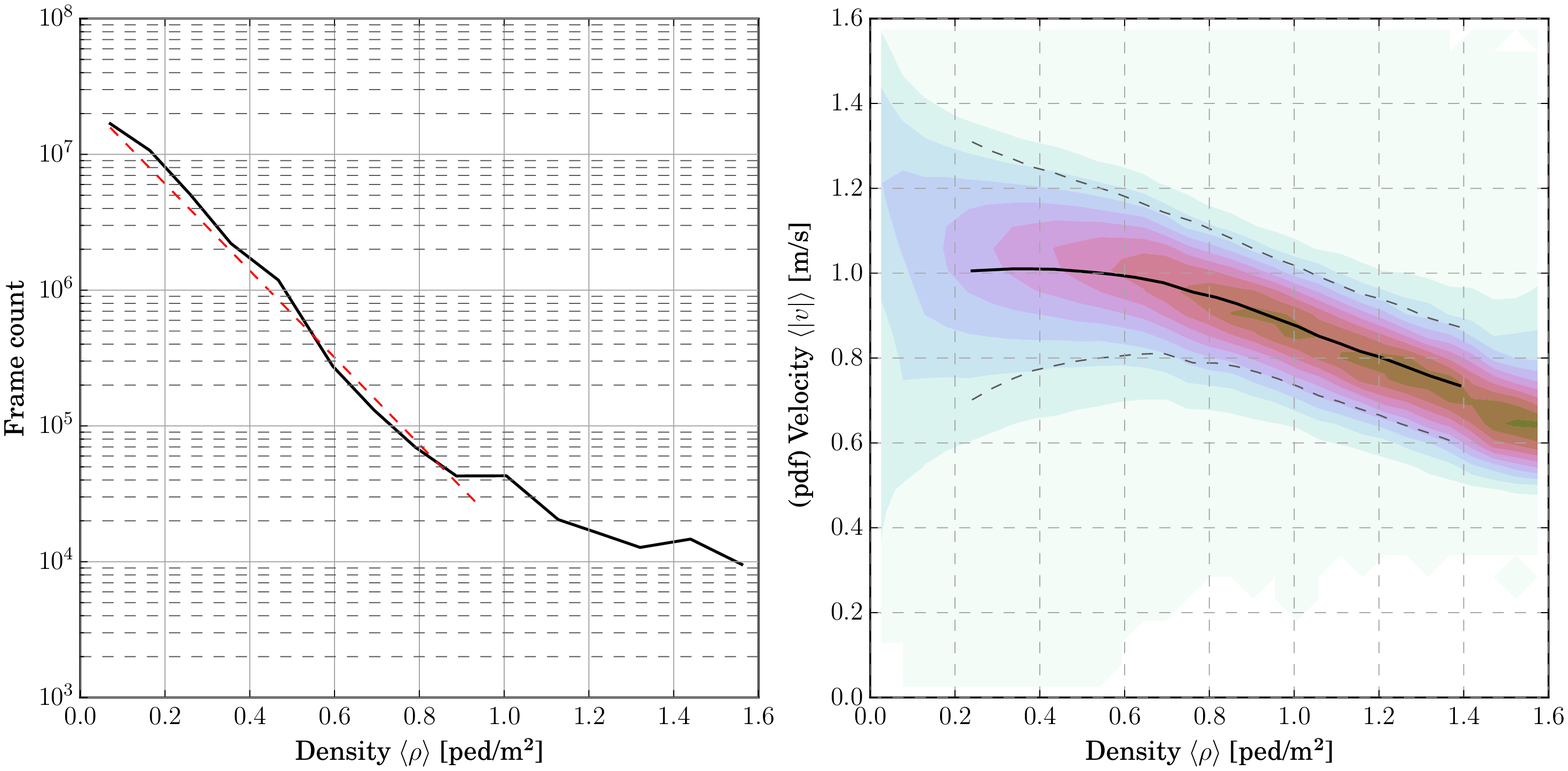}}
\begin{center} 
  \parbox{15.5cm}{\small\center{\bf Fig.\figBasicFdiagr.} (Left)   Count of recorded frames (ca. $38\,$M in total) as a function of the average density $\langle\rho\rangle$. For our installation, and up to $0.95\,$ped/m$^2$, the counts scale exponentially as  $\mbox{count} \approx \mbox{const} \times 10^{-\alpha\langle\rho\rangle}$, with $\alpha \approx 3.2\,$m$^2$/ped (the exponential fitting is plotted as a red dashed line). 
  (Right)    Fundamental diagram: probability density function of (frame-wise) average pedestrian speed $\langle |\rho| \rangle$ vs. average density. High variability is found at low density values as this includes frequent cases of people standing still and running. As the density grows the variability around the average value at each density (solid black line) decreases. The one standard deviation band around the average value is reported via dashed lines.
  }
\end{center}
\vspace*{4mm}

\vspace*{4mm}
\centerline{\includegraphics[width=.40\textwidth,trim=0cm 0cm 0cm 0.cm]{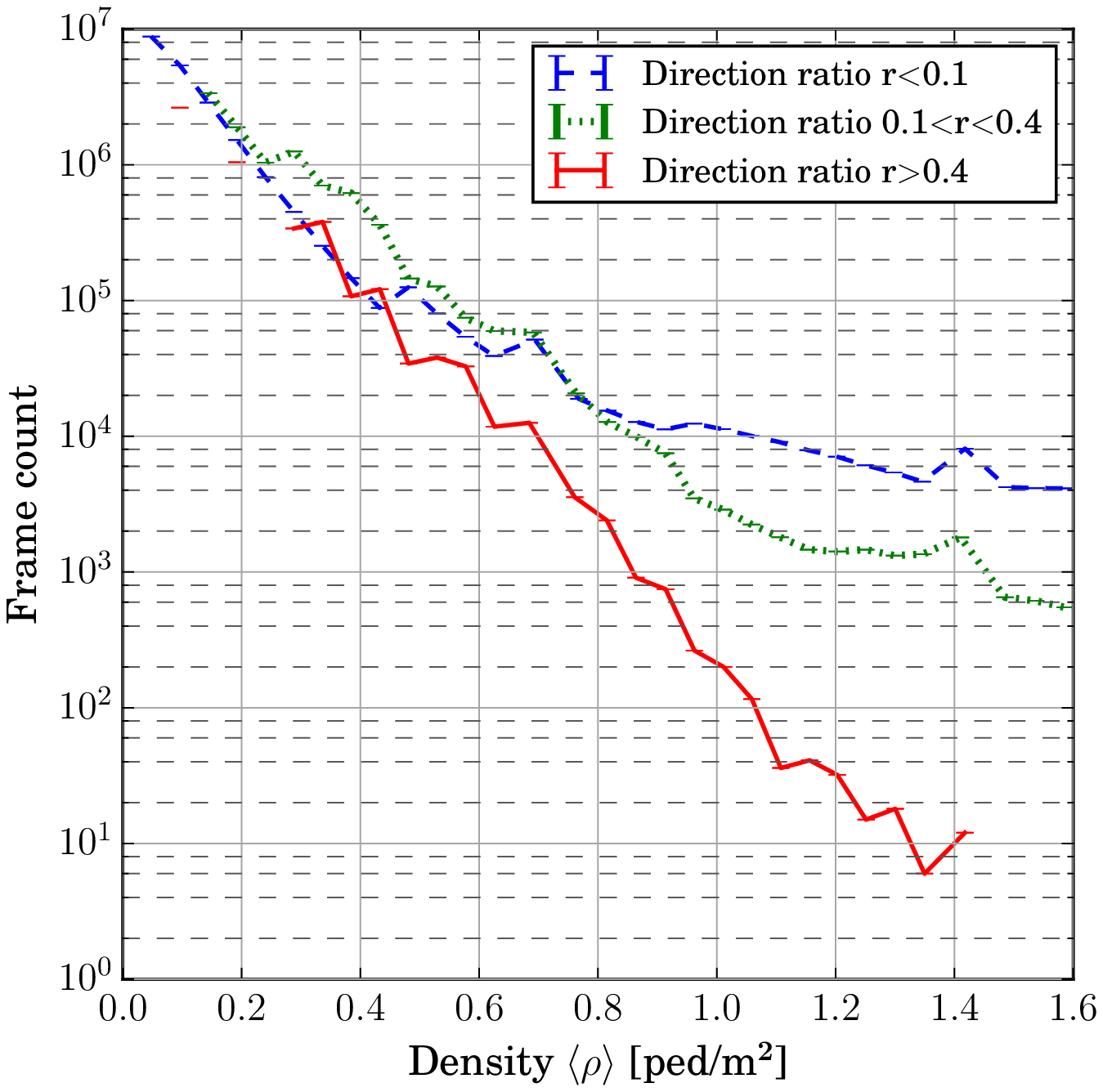}
  \includegraphics[width=.40\textwidth,trim=0cm 0cm 0cm 0.cm]{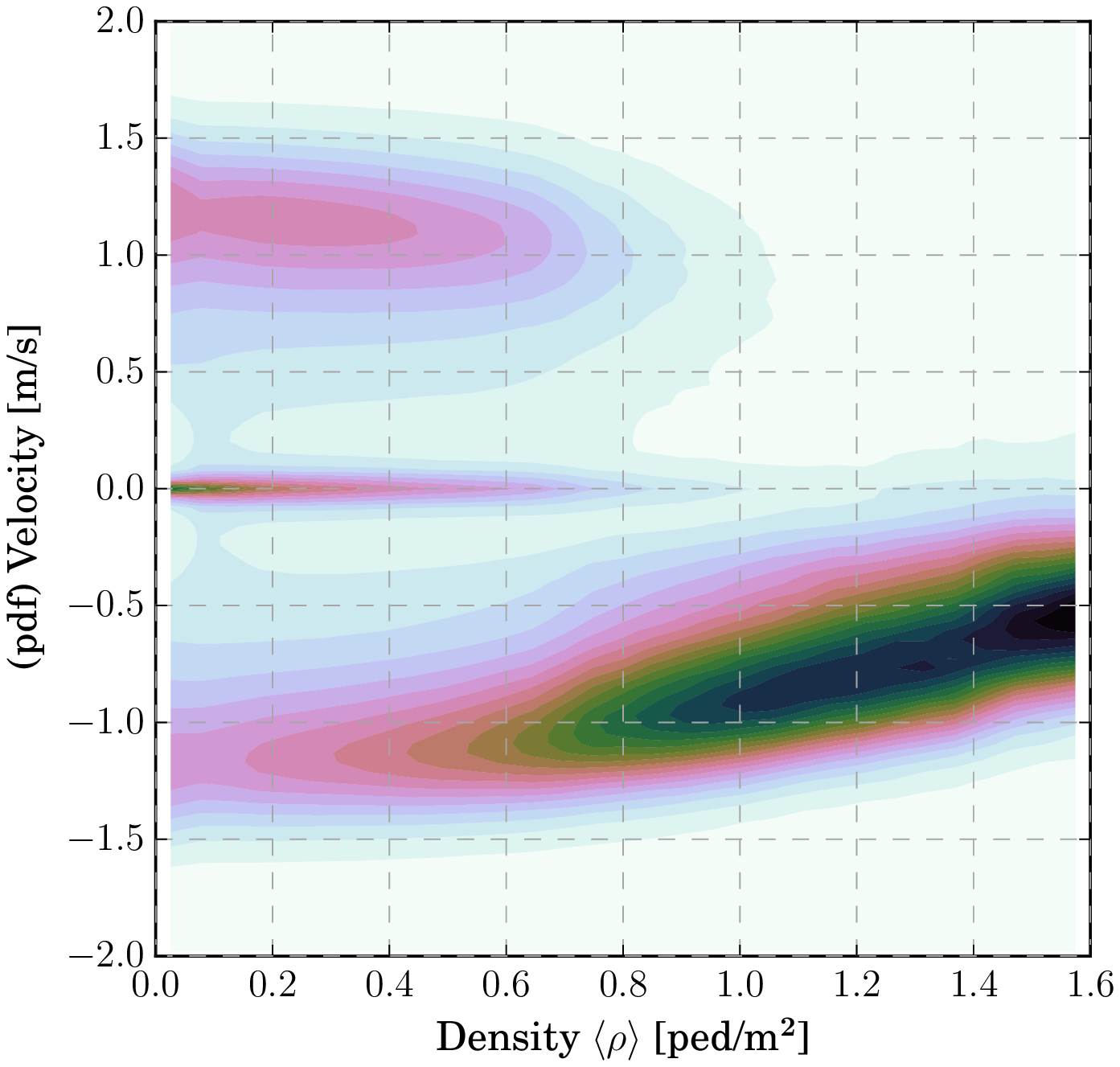}
    }
\begin{center} 
  \parbox{15.5cm}{\small\center{\bf Fig.\figVelRatios.} (Left) Counts of recorded frames vs. density binned by direction ratio $r$ (cf. Eq.~(\ref{eq:direction}) and Fig.~\figSampleTrajs for sample frames at different $r$ values). We bin the measurements in three sets:  $r < 0.1$, i.e. unidirectional flows (dashed line), $0.1<r<0.4$, i.e. uneven bidirectional flows (dotted line), and $r>0.4$, i.e. even bidirectional flows (solid line). (Right) Fundamental diagram: pdf of pedestrian velocity $v$ vs. pedestrian density $\langle \rho \rangle$. In this plot all our velocity measurements contribute equally (i.e. we do not use frame-wise averages). 
 }
\end{center}
\vspace*{4mm}

The fundamental diagram in Fig.~\figBasicFdiagr\,(right)  receives contributions from multiple flow conditions that come in different proportion.  In Fig.~\figVelRatios\,(left) we report the frame counts vs. pedestrian density for different values of the direction ratio $r$ (cf. Eq.~(\ref{eq:direction})). Unidirectional flows ($r<0.1$) retain higher occurrence probabilities in the density intervals $[0,0.2]\,$ped/m$^2$ and $[0.7,1.6]\,$ped/m$^2$. Flows with uneven direction ratios ($0.1 < r < 0.4$) are instead more common in the density range $[0.2,0.7]\,$ped/m$^2$.  Evenly bidirectional flows ($r>0.4$) occur the least, especially at high densities. This is consistent with Fig.~\figVelRatios\,(right), where we report pdfs that include every single velocity measurement $v$ collected in each frame (differently from the frame-wise averages e.g. in Fig.\figBasicFdiagr\,(left)). The velocity $v$ is signed according to the walking direction.  While at low densities we can measure three main dynamics states, namely (i) walking at ca. $1.2\,$m/s toward the center, (ii) walking at ca. $-1.2\,$m/s toward the bus terminal, and (iii) standing still ($0\,$m/s), at high densities negative  velocities dominate (these velocity contributions come from pedestrians moving to the bus terminal, cf. Fig.~\figDensityHistory, hence the fast decay of the $r>0.4$ curve in Fig.~\figVelRatios\,(left)).

\vspace*{4mm}
\centerline{\includegraphics[width=.425\textwidth,trim=0cm 0cm 0cm 0.cm]{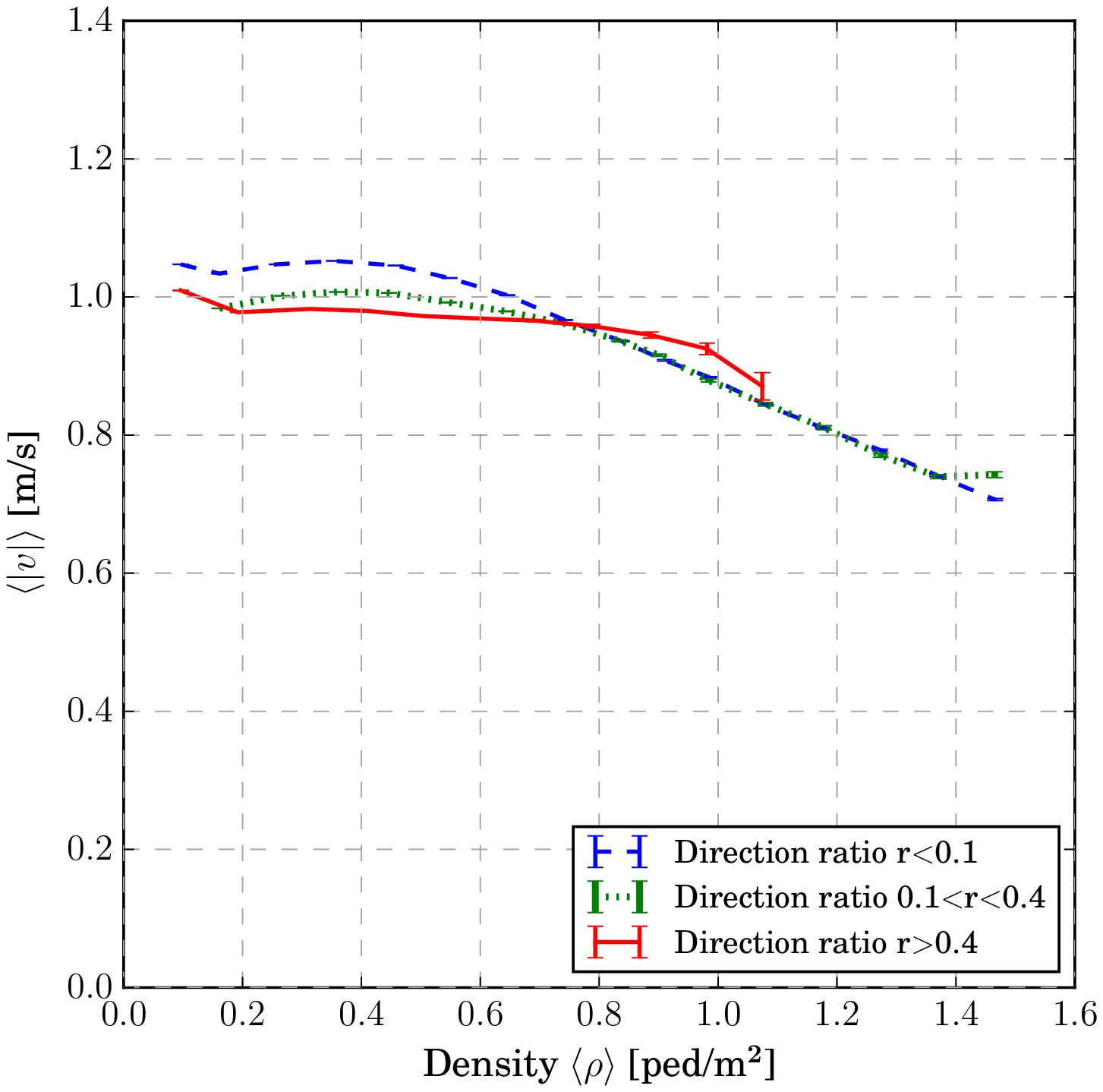}}
\centerline{\includegraphics[width=.85\textwidth,trim=0cm 0cm 0cm 0.cm]{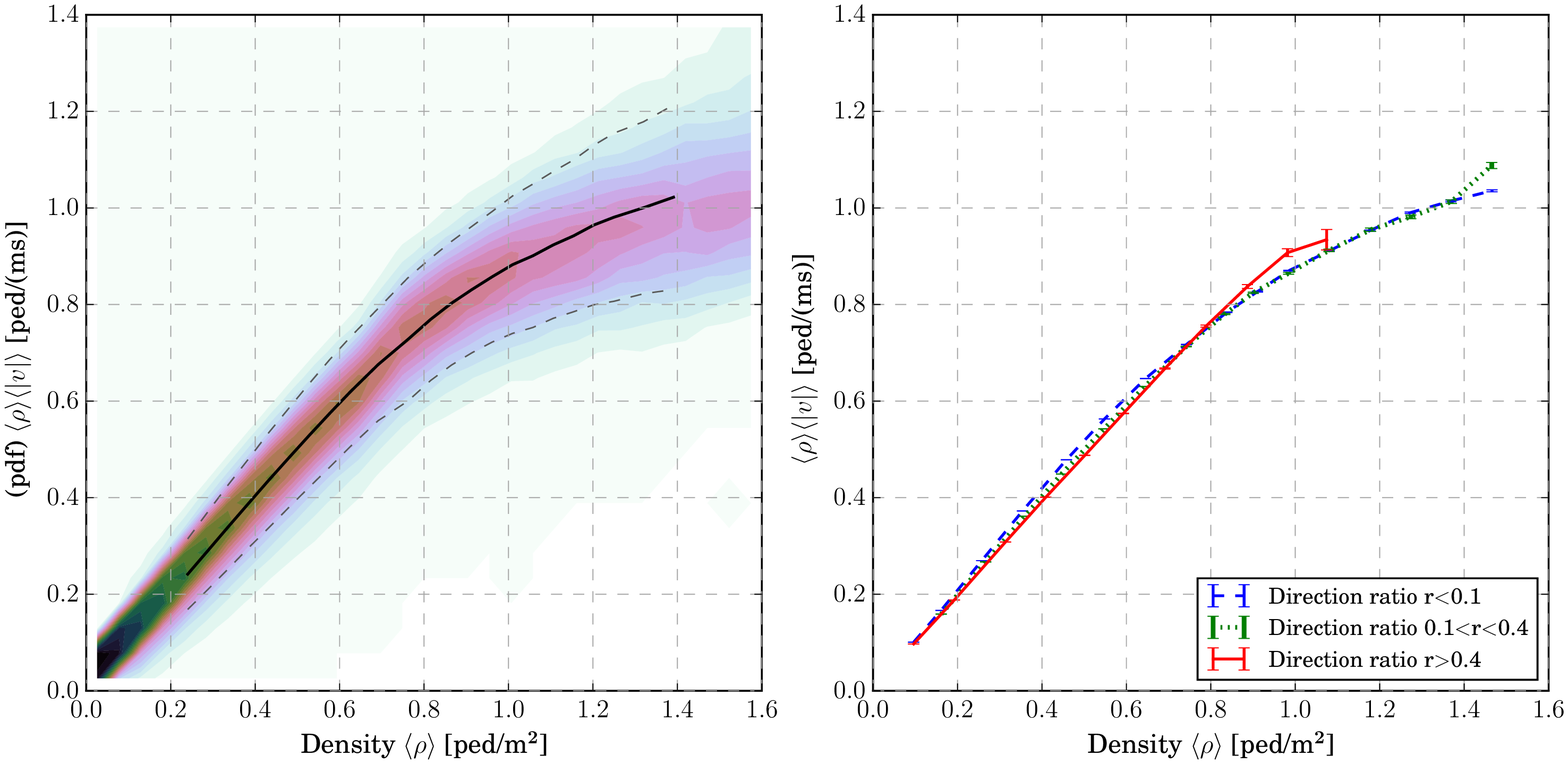}}
\begin{center} 
  \parbox{15.5cm}{\small\center{\bf Fig.\figFluxFdiagr.} (top) Fundamental diagram: average velocity--density relations in dependence on the direction ratios $r$. The same binning on $r$ of Fig.~\figVelRatios\ is used. Error bars report the standard error per density bin ($\sigma/\sqrt{N}$, where $N$ is the total number of measurements in a bin and  $\sigma$ their standard deviation). At densities lower than $0.8\,$ped/m$^2$ unidirectional flow velocities are higher, instead evenly bidirectional flow velocities are higher for $\langle\rho\rangle > 0.8\,$ped/m$^2$. We restrict to flow conditions for which we collected at least $100$ measurements.
    (Left)   Fundamental diagram:  pdf of pedestrian flux ($\langle \rho \rangle\langle |v| \rangle$) vs. pedestrian density ($\langle \rho \rangle$). The average flux is reported with a solid line, while dashed lines depart by plus or minus one standard deviation from the average. These flux measurements include contributions from uni- and bi-directional flows (i.e. at all direction ratios). (Right) Average flux--density relations binned by $r$ as in the top figure.  No significant difference appears in fluxes below $0.8\,$ped/m$^2$, at higher densities a higher flux is measured for evenly bidirectional flows. These conditions occur rarely thus the higher measurement error.}
\end{center}
\vspace*{4mm}

In Fig.~\figFluxFdiagr\,(top) we report frame-wise average velocities $\langle|v|\rangle$ as a function of the density and in dependence on the direction ratio $r$ (these plots synthesize the pdfs in terms of their average value, similarly to the solid line in Fig.~\figBasicFdiagr\,(right)).  To ensure small error in the evaluations, we restrict ourselves to the density-direction ratios combinations for which we collected at least $100$ measurements (cf. Fig.~\figVelRatios\,(left)). Hence, our comparison includes all direction ratios regimes for densities lower than $1.1\,$ped/m$^2$. Data for higher densities are available for unidirectional and unevenly bidirectional flows only. For $\langle\rho\rangle < 0.8\,$ped/m$^2$  unidirectional flow velocities are higher than in any bidirectional regime. At higher densities values, oppositely, we observe higher velocities for even bidirectional flows and no significant difference between velocities from unidirectional and uneven bidirectional flows. Pedestrian fluxes (cf. overall pdf vs. density in Fig.~\figFluxFdiagr\,(left)) are instead almost not distinguishable  for $\langle\rho\rangle < 0.8\,$ped/m$^2$, while evenly bidirectional flux is measured to be the  highest in the interval $[0.8,1.1]\,$ped/m$^2$.

\section{Discussion}\label{sect:concl}
We analyzed the uni- and bi-directional pedestrian dynamics naturally occurring in the $9\,$m wide walkway inside Eindhoven train station. The measurement location is geometrically simple as well as common in terms of pedestrian flows. Our measurements, agglomerating $55$ days of continuous recordings and providing over $38$ millions of non-empty frames, allow an overview of a wide range of dynamics modulated by the periodic arrival of trains and commuters.

We considered universal properties  of the pedestrian flow via  fundamental diagrams, here given in terms of conditioned probability density functions in order to encompass the large data volume. Velocity fundamental diagram show high variability at low densities, including people running and standing still. As the density increases the flow variability decreases.  We further analyzed the pedestrian velocity--density and flux--density relations in dependence on the direction ratio. At densities lower than $0.8\,$ped/m$^2$ we observe higher velocities for unidirectional flows while the trend is inverted between $0.8\,$ped/m$^2$ and $1.1\,$ped/m$^2$ where evenly bidirectional flows show higher velocities and thus fluxes. At densities larger than $1.1\,$ped/m$^2$ we could not collect enough data on the even bidirectional case, the rarest in our measurement site, to enable a comparison. Our measurements are consistent with the idea that bidirectional flows are more efficient than unidirectional flows as the density increases, potentially as a consequence of an increasing self-organization~\cite{kretz,helbing2005self}. Nevertheless, we did not inspect the flow ``degree of organization'' and, in principle, both organized and unorganized dynamics could have occurred and been agglomerated in our statistics. A detailed analysis in this direction is left for future work.

Finally we notice that we operated frame-wise analyses, i.e. Eulerian in the sense of~\cite{corbetta2016eulerian}, thus based on instantaneous measurements. We leave for future work the inclusion of trajectory-wise properties (Lagrangian), such as trajectory-wise velocity averages, trajectory-wise personal spaces \textit{et cetera}, that we expect to further enrich our insights on the dynamics of pedestrians.

\section*{Acknowledgements}
We acknowledge the Brilliant Streets research program of the
Intelligent Lighting Institute at the Eindhoven University of
Technology, Nederlandse Spoorwegen, and the technical support of A. Muntean, T. Kanters, A. Holten, G. Oerlemans, and M. Speldenbrink. During the development  of the infrastructure, AC has been partly founded by a Lagrange Ph.D. scholarship granted by 
the CRT Foundation, Turin, Italy and by the Eindhoven University
of Technology, the Netherlands. This work is part of the JSTP research programme ``Vision driven visitor behaviour analysis and crowd management" with project number 341-10-001, which is financed by the Netherlands Organisation for Scientific Research (NWO).
  
\bibliographystyle{unsrt}
\bibliography{bibliog}

\begin{thebibliography}{10}

\bibitem{DBLP:journals/ijon/BoltesS13}
M.~Boltes and A.~Seyfried.
\newblock Collecting pedestrian trajectories.
\newblock {\em Neurocomputing}, 100:127--133, 2013.

\bibitem{seer2014kinects}
S.~Seer, N.~Br{\"a}ndle, and C.~Ratti.
\newblock Kinects and human kinetics: A new approach for studying pedestrian
  behavior.
\newblock {\em Transport. Res. C-Emer.}, 48:212--228, 2014.

\bibitem{Kinect}
{Microsoft Corp.}
\newblock Kinect for {X}box 360, available online:
  http://www.xbox.com/en-us/kinect/, 2011.
\newblock Redmond, WA, USA.

\bibitem{Brscic201477}
D.~Br\v{s}\v{c}i\'{c}, F.~Zanlungo, and T.~Kanda.
\newblock Density and velocity patterns during one year of pedestrian tracking.
\newblock {\em Transportation Research Procedia}, 2:77 -- 86, 2014.

\bibitem{corbetta2014TRP}
A.~Corbetta, L.~Bruno, A.~Muntean, and F.~Toschi.
\newblock High statistics measurements of pedestrian dynamics.
\newblock {\em Transportation Research Procedia}, 2:96--104, 2014.

\bibitem{corbetta2016eulerian}
A.~Corbetta, C.~Lee, A.~Muntean, and F.~Toschi.
\newblock Eulerian vs. lagrangian analyses of pedestrian dynamics asymmetries
  in a staircase landing.
\newblock {\em arXiv preprint arXiv:1606.03946}, 2016.

\bibitem{corbetta2016multiscale}
A.~Corbetta.
\newblock {\em Multiscale crowd dynamics: physical analysis, modeling and
  applications}.
\newblock PhD thesis, Technische Universiteit Eindhoven, 2016.

\bibitem{weid}
U.~Weidmann.
\newblock Transporttechnik der {F}ussg\"{a}nger.
\newblock Technical Report~90, ETH, Z\"{u}rich, 1993.

\bibitem{fruin1987BOOK}
J.~J. Fruin.
\newblock {\em Pedestrian Planning and Design}.
\newblock {E}levator {W}orld {I}nc., 1987.

\bibitem{pushkarev1975capacity}
B.~Pushkarev and J.M. Zupan.
\newblock Capacity of walkways.
\newblock {\em Transport. Res. Rec.}, 538:1--15, 1975.

\bibitem{lam2002study}
W.~Lam, J.~Lee, and C.~Cheung.
\newblock A study of the bi-directional pedestrian flow characteristics at hong
  kong signalized crosswalk facilities.
\newblock {\em Transportation}, 29(2):169--192, 2002.

\bibitem{zhang2012ordering}
J.~Zhang, W.~Klingsch, A.~Schadschneider, and A.~Seyfried.
\newblock Ordering in bidirectional pedestrian flows and its influence on the
  fundamental diagram.
\newblock {\em J. Stat. Mech.-Theory E.}, 2012(02):P02002, 2012.

\bibitem{kretz}
T.~Kretz, A.~Gr\"{u}nebohm, M.~Kaufman, F.~Mazur, and M.~Schreckenberg.
\newblock Experimental study of pedestrian counterflow in a corridor.
\newblock {\em J. Stat. Mech.-Theory E.}, 2006(10):10001, 2006.

\bibitem{helbing2005self}
D.~Helbing, L.~Buzna, A.~Johansson, and T.~Werner.
\newblock Self-organized pedestrian crowd dynamics: Experiments, simulations,
  and design solutions.
\newblock {\em Transport. Sci.}, 39(1):1--24, 2005.

\bibitem{corbettaTGF15}
A.~Corbetta, C.~Lee, A.~Muntean, and F.~Toschi.
\newblock Asymmetric pedestrian dynamics on a staircase landing from continuous
  measurements.
\newblock In W.~Daamen and V.~Knoop, editors, {\em Traffic and Granular Flows
  '15}, chapter~7. Springer, 2016.

\bibitem{corbetta2015MBE}
A.~Corbetta, A.~Muntean, and K.~Vafayi.
\newblock Parameter estimation of social forces in pedestrian dynamics models
  via a probabilistic method.
\newblock {\em Math. Biosci. Eng.}, 12(2):337--356, 2015.

\bibitem{brscic2013person}
D.~Br\v{s}\v{c}i\'{c}, T.~Kanda, T.~Ikeda, and T.~Miyashita.
\newblock Person tracking in large public spaces using 3-d range sensors.
\newblock {\em IEEE Trans. Human-Mach. Syst.}, 43(6):522--534, 2013.

\bibitem{OpenPTV}
{The {OpenPTV} Consortium}.
\newblock {OpenPTV: Open source particle tracking velocimetry, available
  online: http://www.openptv.net/}, since 2012.

\bibitem{willneff2003spatio}
J.~Willneff.
\newblock {\em A Spatio-temporal Matching Algorithm for 3D Particle Tracking
  Velocimetry}.
\newblock PhD thesis, ETH Z{\"u}rich, 2003.

\bibitem{steffen2010methods}
B.~Steffen and A.~Seyfried.
\newblock Methods for measuring pedestrian density, flow, speed and direction
  with minimal scatter.
\newblock {\em Physica A}, 389(9):1902--1910, 2010.

\end{thebibliography}

\end{document}